\documentclass[sigconf]{acmart}
\acmConference[CAIN 2024]{3rd International Conference on AI Engineering — Software Engineering for AI}{April 2024}{Lisbon, Portugal}
\usepackage{graphicx} %
\usepackage{booktabs, multirow} %
\usepackage{soul}%
\usepackage{changepage,threeparttable} %
\usepackage{colortbl}
\usepackage{hhline}

\title[Energy Consumption in DL Models Across Different Runtime Infrastructures]{Green AI: A Preliminary Empirical Study on Energy Consumption in DL Models Across Different Runtime Infrastructures}
\author{Negar Alizadeh}
\email{n.s.alizadeh@uu.nl}
\affiliation{%
\institution{Utrecht University \country{The Netherlands}}
}
\author{Fernando Castor}
\email{f.castor@utwente.nl}
\affiliation{%
\institution{University of Twente \country{The Netherlands}}
\institution{Utrecht University \country{The Netherlands}}
\institution{Universidade Federal de Pernambuco \country{Brazil}}
}
\date{}

\begin{document}
\begin{abstract}
Deep Learning (DL) frameworks such as PyTorch and TensorFlow include runtime infrastructures responsible for executing trained models on target hardware, managing memory, data transfers, and multi-accelerator execution, if applicable. 
Additionally, it is a common practice to deploy pre-trained models on environments distinct from their native development settings. This led to the introduction of interchange formats such as ONNX, which includes its runtime infrastructure, and ONNX Runtime, which work as standard formats that can be used across diverse DL frameworks and languages. Even though these runtime infrastructures have a great impact on inference performance, no previous paper has investigated their energy efficiency.
In this study, we monitor the energy consumption and inference time in the runtime infrastructures of three well-known DL frameworks as well as ONNX, using three various DL models. To have nuance in our investigation, we also examine the impact of using different execution providers.
We find out that the performance and energy efficiency of DL are difficult to predict. One framework, MXNet, outperforms both PyTorch and TensorFlow for the computer vision models using batch size 1, due to efficient GPU usage and thus low CPU usage. However, batch size 64 makes PyTorch and MXNet practically indistinguishable, while TensorFlow is outperformed consistently. For BERT, PyTorch exhibits the best performance. Converting the models to ONNX yields significant performance improvements in the majority of cases.  Finally, in our preliminary investigation of execution providers, we observe that TensorRT always outperforms CUDA.

\end{abstract}


\maketitle

\section{Introduction}
Although DL plays a crucial role in streamlining and simplifying many activities, its usage is extremely energy-intensive \cite{georgiou2022green, garcia2019estimation}. Apart from its massive financial cost, it incurs high carbon emissions~\cite{strubell2019energy,lacoste2019quantifying} and as a result, contributes to climate change. Historically, the literature on DL has focused mostly on improving model accuracy. Notwithstanding, the widespread use of this technology combined with its intensive resource usage and the growing threat of climate change demands a shift in priorities, so as to make it greener~\cite{schwartz2019green}.

Developers face multiple technical decisions impacting DL-based solution outcomes, including framework, optimizer, architecture, batch size, and runtime infrastructure choices that influence the model's performance and energy footprint. Recent papers recognize this and study the impact of different elements such as frameworks~\cite{georgiou2022green, openja2022empirical}, batch sizes~\cite{yarally2023batching}, and DL architectures~\cite{Yarally:2023:UEE}, considering both the training~\cite{georgiou2022green, openja2022empirical,Yarally:2023:UEE} and inference~\cite{georgiou2022green, openja2022empirical,yarally2023batching} phases. 

The runtime infrastructure of a DL framework refers to the set of software components and libraries that enable the execution and deployment of DL models, i.e., for the inference phase. This includes the environment in which trained models run, make predictions, and interact with data. Popular DL frameworks e.g. TensorFlow and PyTorch provide runtime infrastructure components for deploying and running DL models. Since this is a performance-critical element of a DL pipeline, there are specialized runtime infrastructures e.g. ONNX~\cite{LF:2019} that aim to improve the efficiency of inference for pre-trained models. Previous work~\cite{openja2022empirical} investigated the performance of ONNX in terms of execution time and memory usage for models trained on PyTorch and TensorFlow. 

In this paper, we contribute to the existing body of knowledge by examining the energy usage of running inference on ONNX, versus the runtime infrastructures of three DL frameworks, PyTorch, TensorFlow, and MXNet. Furthermore, we study how ONNX's performance is influenced by the selection of different execution providers and back-end libraries for execution on GPUs. More specifically, we target two such execution providers: CUDA~\cite{nvidia:2007} and TensorRT~\cite{nvidia:2021}. We measure the energy usage of both CPU and GPU, as well as the accuracy and execution time of inference (inference time) for three different models, ResNet, MobileNet, and BERT-base-uncased. To the best of our knowledge, this is the first paper investigating the energy footprint of ONNX and also the first to study how different execution providers affect its inference performance.

We find out that, on one hand, inference time and energy consumption exhibit a very strong correlation (spearman's rho=0.99, p $<$ 0.0001) when considering the aggregated results of the study. 
On the other hand, the performance and energy efficiency of DL are fickle and difficult to predict. MXNet outperforms both PyTorch and TensorFlow for vision models in batch size 1, due to efficient GPU usage and thus low CPU usage. However, batch size 64 makes PyTorch and MXNet practically indistinguishable, while TensorFlow is outperformed consistently. For BERT, PyTorch outperforms the other frameworks and MXNet is surprisingly inefficient with batch size 1. 
Converting the models to ONNX usually yields significant performance improvements with a few exceptions for ResNet and Mobilenet with batch size 64. In these cases, converted models of MXNet and PyTorch consume around 10-13\% more energy and time than the original models.
Finally, in our preliminary investigation of execution providers, TensorRT always outperforms CUDA.

\section{Related Work}
This section presents a comprehensive review of prior research dedicated to examining energy efficiency in software applications, with a specific focus on Machine Learning (ML). Although the study of energy in software systems is a relatively mature field \cite{duarte2019model,Pinto:2017:EEN,stoico2023approach}, as it was represented in \cite{watson2022systematic}, there were not many studies concerning the energy consumption of ML in software applications. The majority of the studies examined in this review are recent publications. This highlights a growing emphasis on understanding the importance of energy usage in this particular domain.

While some studies have paved the way in \textbf{Green AI} \cite{schwartz2019green, yarally2023uncovering, georgiou2022green, desislavov2021compute}, there are few studies focusing specifically on the Inference phase \cite{desislavov2021compute}. Desislavov et al. highlighted the significance of inference costs, 
an aspect often overshadowed in research papers, where the focus typically leans toward training costs.
The study exposes that, given the recurrent utilization of numerous DL models (trained only once but applied millions of times), the cumulative impact is notably higher.
Xu et al. \cite{xu2023energy}
explored the impact of different CNN architectures on the computer vision domain on the energy efficiency of the model training phase, as well as the likely relation between energy efficiency and the accuracy of the obtained model.
Additionally, \cite{omar2023ai, verdecchia2022data} took a data-centric approach, and shed a spotlight on the potential effects of modifying datasets to enhance the energy usage of training AI models. Fischer et al. \cite{fischer2023energy} also investigated the energy efficiency of deploying different ML methods on popular classification benchmark datasets from Scikit-learn and OpenML. Gutiérrez et al. \cite{gutierrez2022analysing} studied the energy impact of various optimization methods of Logistic Regression in a fraud detection application, only to reveal negligible accuracy gains after employing some solvers at the expense of significantly more energy cost.

Henderson et al. \cite{henderson2020towards} proposed a framework to provide insight into energy consumption and carbon emissions of ML systems and utilized it to investigate the energy efficiency of reinforcement learning algorithms. Similarly, Anthony et al. \cite{anthony2020carbontracker} proposed a tool for tracking and predicting the energy consumption and carbon emissions of training DL models in different environments and platforms. In a more fine-grained analysis, Rajput et al. \cite{rajput2023fecom} presented a framework that measures consumed energy at the method level.

Openja et al. \cite{openja2022empirical} presented an empirical study assessing the effectiveness of conversion formats, ONNX and CoreML, in the accuracy, performance, and robustness of pre-trained DL models. The study used Keras and PyTorch to train five widely used models across three datasets. Results indicate that the accuracy is the same while converted models have smaller sizes. Similarly, 
in a very recent study, Jacques et al. \cite{Jacques2024battery} compared the inference time and battery usage of DL models and their converted versions on iOS devices. Georgiou et al. \cite{georgiou2022green} conducted experiments on PyTorch and TensorFlow in both training and inference, aiming to gauge the energy efficiency and performance of each one. They further investigated the trade-off between resource usage and accuracy. While the methodologies of the three recent studies closely align with ours, It is worth mentioning that our investigation encompasses a different scope. We include MXNet in addition to TensorFlow and PyTorch, and for the ONNX converted models, we delve deeper by examining its performance on two different execution providers: CUDA and TensorRT. Furthermore, it's worth highlighting that energy considerations were not accounted for in the first study~\cite{openja2022empirical}.

\section{Methodology}\label{sec:meth}
In this section, we describe our benchmarking procedure and provide further detail on the frameworks under consideration, the DL models under examination, and the experimental setup as well as the methodology employed in our experiments. Fig. 1 shows an overall diagram of our proposed approach, detailed in the following paragraphs.
We set out to answer these three research questions:
\begin{description}
\item[RQ1] How do the frameworks differ in terms of energy based on the models under examination?
\item[RQ2] How is a converted pre-trained model (ONNX version) different from its originally implemented equivalent model in terms of energy, performance, and accuracy?
\item[RQ3] How do the execution providers affect the performance of inference for models running on ONNX?

\end{description}

\vspace{0.1cm}
\noindent
\textbf{Frameworks.}
In this study, we investigate three DL frameworks, PyTorch (v1.12.1), TensorFlow/Keras (v2.9.1), and MXNet (v1.9.1). 
PyTorch and TensorFlow are the two most popular DL frameworks and MXNet is recognized for its efficiency and scalability,
which makes it a good choice for energy-centered studies. It is supported by cloud providers such as Microsoft's Azure and Amazon's AWS. In particular, Amazon has selected it as its DL framework of choice, due to its portability, development speed, and ability to scale~\cite{Vogels:2016:MDL}.

\begin{figure}[tb]
\centerline{
\includegraphics[width=\linewidth]
{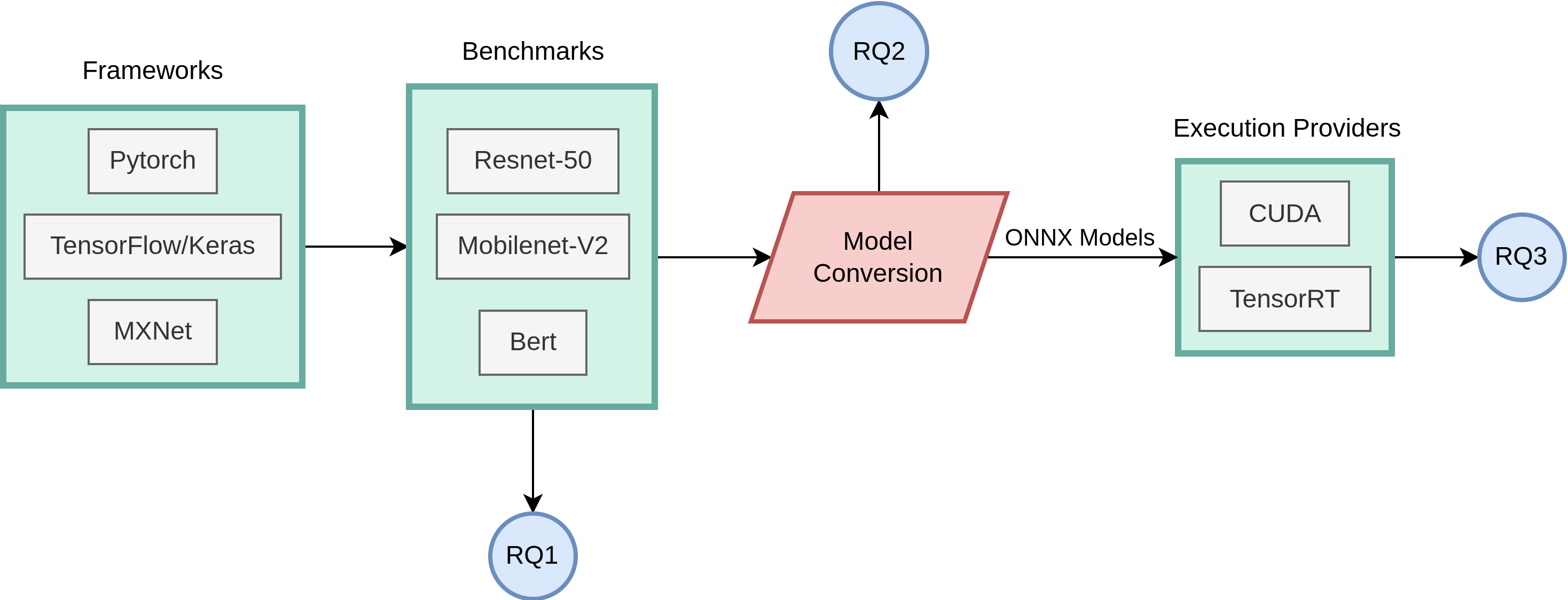}}
\caption{Schematic representation of this study.}
\label{fig1}
\end{figure}

\vspace{0.15cm}
\noindent
\textbf{Deep Learning Models.}
The studied benchmarks are categorized into two main groups: Image Classification (Computer Vision) and Text Classification (Sentiment Analysis). Each category includes pre-trained models implemented using the mentioned frameworks, all of which are official pre-trained models available in their respective repositories. Since these models are trained and distributed by the framework developers, we assume that they are best-of-breed solutions, and comparing them is, as a consequence, fair.

For image classification, we investigate 2 different models.
\textit{ResNet-50} is a CNN architecture consisting of 50 layers. 
ResNet and its variants have had a great impact on the field of DL, serving as a basis for many state-of-the-art models \cite{he2016deep}.
\textit{MobileNetV2} 
is known for its efficiency and suitability for deployment on devices with less computational power. This characteristic makes it a good candidate for such studies focused on energy efficiency \cite{howard2017mobilenets}.
Both ResNet-50 and MobileNetV2 are pre-trained (official) models on the ImageNet-1k dataset, a large dataset containing images from 1000 categories. 

For text classification, we study Bert-base-uncased, a transformer-based language model that can be fine-tuned to create state-of-the-art models for a wide range of tasks  \cite{devlin2018bert}.
In this study, it is fine-tuned on the IMDB reviews dataset to be used as a text classifier.

\vspace{0.15cm}
\noindent
\textbf{Methodology for RQ1.} To address RQ1, we conduct pairwise comparisons of the mentioned models across various frameworks. Initially, for the vision tasks, we chose a subset of 10000 images from the ImageNet dataset. 
For the BERT model, we choose 25000 reviews from the IMDB dataset and use it to predict the sentiment polarity of each one. We consider two different batch sizes, 1 and 64. The first one would be a typical scenario of the model being used to classify a single image. The second one simulates an alternative scenario involving the offline inference of a dataset containing the whole 10000 images. In both cases, we preprocess all images to meet the specific requirements of each model and then supply ready-to-use input images to the models. We record the accuracy and inference time as well as the energy consumption.

\vspace{0.15cm}
\noindent
\textbf{Methodology for RQ2.} For this research question, we include the ONNX Runtime (v1.14) in the comparison. Since ONNX is an interchange format, it is possible to export models trained on different DL frameworks to ONNX format. We use MXNet, PyTorch, and TensorFlow built-in support for ONNX to export models trained in each of these frameworks. These models can then be optimized and deployed on various hardware platforms. We execute the ONNX models similarly to what we do for RQ1 in the case of ResNet and MobileNet. For BERT, the models exported by MXNet and TensorFlow were considered invalid by the ONNX Runtime. Therefore, we only report results for the model exported by PyTorch. All the experiments for this RQ were conducted using CUDA as the ONNX execution provider (back-end). 

\vspace{0.15cm}
\noindent
\textbf{Methodology for RQ3.} The ONNX Runtime can be configured with different execution providers to better leverage the computing power of GPUs. These execution providers are libraries that act as the back-end of the ONNX Runtime, optimizing execution for GPUs. In this work, we examine two execution providers: NVIDIA CUDA and NVIDIA TensorRT. CUDA is the most popular library to express computations for NVIDIA GPUs. TensorRT aims to support DL on NVIDIA GPUs by implementing multiple optimizations for DL inference, e.g. quantization and layer and tensor fusion. For this comparison, we focus solely on ResNet after conversion from each framework to ONNX. To conduct our experiments, we utilized an NVIDIA NGC Docker container (Pytorch 22.03)\cite{nvidia-ngc:2023}, specifically optimized for GPU acceleration. This container incorporates a validated set of libraries designed to enable and optimize GPU performance, including essential components such as cuDNN and TensorRT, necessary for our experiments.

\vspace{0.15cm}
\noindent
\textbf{Execution Environment.} All these experiments have been conducted on a laptop with Ubuntu desktop 20.04-64bit and equipped with an NVIDIA Geforce RTX 3070 GPU with 8 GB memory and an Intel Core i7-11850H CPU, 2.5 GHz frequency with 31 GB memory and 24MB cache.
To capture GPU power usage and utilization, we set the GPU to high-performance mode. 
We query the NVIDIA System Management Interface (nvidia-smi)\cite{nvidia:2023} every 10 milliseconds. The total energy consumption can then be computed as a factor of time and the average power \cite{yarally2023batching}. To normalize the GPU power usage, we record the GPU power in idle mode for 10 minutes and subtract the average from our result. As for the energy consumption of the CPU we employ pyRAPL \cite{PyRAPL:2023}, a Python library designed for interacting with the Running Average Power Limit (RAPL) interface. RAPL is available for Intel processors and has been widely utilized in previous research studies ~\cite{oliveira2021improving,georgiou2022green}, and has accuracy comparable to that of using external measurement equipment~\cite{Khan:2018:RAE}. Similarly, for establishing a baseline, we gauge CPU energy consumption during its idle state for 10 minutes, then compute the average power, and subtract it from our recorded results.
\section{Results}
This section presents the outcomes of our analysis carried out in the previous step for each research question. Tables~\ref{tab:rq1}, \ref{tab:framework_accuracies}, and~\ref{tab:rq3} summarize the main results of our study. These findings not only contribute to the existing body of knowledge in enhancing awareness regarding energy consumption but also have implications for energy optimization. Each subsection in this section addresses one of the research questions.
\begin{table*}[tb]
\centering
\label{rq1}
\caption{Comparative results of inference in MobilenetV2, Resnet-50, Bert-base-uncased across Pytorch, TF/Keras, and MXNet, including insights on ONNX versions.}\label{tab:rq1}
\resizebox{\textwidth}{!}{%
\begin{tabular}{lclrrrr>{\columncolor[HTML]{D3F3E7}}r>{\columncolor[HTML]{DAE8FC}}r}
\hline
Model &
\begin{tabular}[r]{@{}c@{}}Batch\\size\end{tabular} &
  Framework &
  \begin{tabular}[r]{@{}r@{}}GPU Utilization (\%)\\ (Cores, Memory)\end{tabular} &
  GPU Power (W) &
  GPU Energy (J) &
  CPU Energy (J) &
  Total Energy (J) &
  \begin{tabular}[r]{@{}r@{}}Inference\\ Time (Sec)\end{tabular}
   \\ \hline
\multirow{12}{*}{MobileNet} & \multirow{6}{*}{1}  & mxnet      & (68, 13.16) & 42.18  & 956.95  & \textbf{566.31}  & \textbf{1523.27} & \textbf{22.69}       \\
                  &   & tf         & (7.97, 1.02) & 2.71  & 610.5   & 4457.76 & 5068.26  & 225.3                               \\
                  &   & torch        & (32.27, 6.11) & 12.3  & \textbf{465.88}  & 1136    & 1601.88  & 37.86                               \\ 
                  \hhline{~~-------}
                  &   & mx2onnx    & (87.03, 16.27) & 50.88  & 830.97  & 258.83  & 1089.8 & 16.33                              \\
                  &   & tf2onnx    & (83.86, 15.56) & 48.42  & 809.09  & 259.99  & 1069.08 & 16.71                              \\
                  &   & torch2onnx & (82.55, 13.25) & 47.52  & \textbf{771.22}  & \textbf{247.94}  & \textbf{1019.15} & \textbf{16.23}                                \\ \hhline{~--------}
& \multirow{6}{*}{64} & mxnet      & (68.15, 63.17) & 38.58   & \textbf{341.48}  & 222.73  & 564.21 & 8.85                                 \\
                  &   & tf         & (51.05, 43.32) & 38.9  & 568.82  & 281.8   & 850.62  & 14.62                                 \\
                  &   & torch      & (70.38, 61.99) & 42.75    & 363.24  & \textbf{179.67}  & \textbf{542.91} & \textbf{8.5}                                 \\ \hhline{~~-------}
                  &   & mx2onnx    & (80.62, 68.9) & 47.6  & 477.54  & 158.6   & 636.14  & 10.03                                  \\
                  &   & tf2onnx    & (81.64, 70.7) & 48.16  & 492.56  & 168.81  & 661.36  & 10.23                                 \\
                  &   & torch2onnx & (80.83, 69.53) & 47.35   & \textbf{444.67}  & \textbf{150.64}  & \textbf{595.31} & \textbf{9.39}                                 \\ \hline
\multirow{12}{*}{ResNet}  & \multirow{6}{*}{1}  & mxnet      & (74.08, 23.28) & 50.82  & 2586.06 & \textbf{847.9}   & \textbf{3433.96} & \textbf{50.89}      \\
                  &   & tf         & (18.58, 6.37) & 10.79 & 2915.05 & 5315.09 & 8230.14 & 270.09                                 \\
                   &  & torch      & (51.67, 21.65) & 49.1  & \textbf{2573.77} & 1431.36 & 4005.14  & 52.42                                \\ \hhline{~~-------}
                  &   & mx2onnx    & (94.06, 30.68) & 53.17  & 2269.03 & \textbf{541.27}  & 2810.3 & 42.67                                 \\
                  &   & tf2onnx    & (90.57, 26.89) & 53.02  & 2057.46 & 593.3   & \textbf{2650.77} & 38.81                                \\
                  &   & torch2onnx & (94.25, 25.86) & 53.01  & \textbf{2039.06} & 628.67  & 2667.73 & \textbf{38.46}                                \\ \hhline{~--------}
& \multirow{6}{*}{64} & mxnet      & (85.3, 65.45) & 47.1  & \textbf{914.6}   & 358.76  & \textbf{1273.35}  & 19.42                                 \\
                  &   & tf         & (62.62, 50.25) & 44.78  & 1230.91 & 492.31  & 1723.22 & 27.49                                \\
                  &   & torch      & (86.61, 66.43) & 49.45  & 937.06  & \textbf{351.99}  & 1289.05 & \textbf{18.95}                                \\ \hhline{~~-------}
                  &   & mx2onnx    & (91.11, 74.63) & 51.81   & 1114.02 & 296.08  & 1410.1 & 21.5                                 \\
                  &   & tf2onnx    & (78.63, 52.51) & 50.63  & 811.94  & 219.5   & 1031.44 & 16.04                                \\
                  &   & torch2onnx & (88.27, 54.77) & 50.48  & \textbf{792}     & \textbf{219.35}  & \textbf{1011.35} & \textbf{15.69}                                \\ \hline 
\multirow{8}{*}{BERT} & \multirow{4}{*}{1} & mxnet & (38.02, 16.89) & 25.84 & 8547.11 & 6632.06 & 15179.17 & 330.78   \\
                  &   & tf         & (75.96, 21.32) & 39.37 & \textbf{7963.21} & 3566.74 & 11529.95 & 202.26                               \\
                  &   & torch      & (47.91, 22.52) & 51.78 & 8243.54 & \textbf{1648.22} & \textbf{9891.76} & \textbf{159.21}                                \\ \hhline{~--------}
                  &   & torch2onnx & (96.77, 42.94) & 54.26  & \textbf{3904.3}  & \textbf{845.93}  & \textbf{4750.24} & \textbf{71.96}                                \\ \hhline{~--------}
& \multirow{4}{*}{64} & mxnet      & (83.55, 46.09) & 53.42  & 4227.29 & 1261.12 & 5488.41 & 79.13                                \\
                  &   & tf         & (80.86, 53.32) & 47.24  & 6718.37 & 2130.84 & 8849.21 & 142.2                                \\
                  &   & torch      & (96.23, 35.21) & 53.36  & \textbf{2250.15} & \textbf{630.38}  & \textbf{2880.53} & \textbf{42.17}                                \\ \hhline{~--------}
                  &   & torch2onnx & (\textbf{99.32}, 46.49) & 54.16  & \textbf{1362.62} & \textbf{404.17}  & \textbf{1766.79} & \textbf{25.16}                                \\\hline
\end{tabular}%
}
\end{table*}
\begin{table}[tb]
\label{rq2}
\centering
{\small
\caption{Framework x Model accuracy}
\label{tab:framework_accuracies}
\begin{tabular}{|l|c|c|c|}
\hline
Framework & MobileNetV2 & ResNet50 & BERT \\
          & ImageNet    & ImageNet & IMDB \\
\hline
MXNet & 69.9 & 75.9 & \textbf{88.42} \\
TF/Keras & \textbf{71.07} & 70.94 & 77.21 \\
Pytorch & 70.48 & \textbf{80.17} & 76.78 \\
\hline
\end{tabular}
}
\end{table}
In terms of general trends, we can confirm the folklore: total time and total energy have a very strong correlation in the context of DL frameworks (spearman's rho=0.99, pvalue=6.54e-27). This contrasts a bit with some other results on the relationship between these two variables, in the context of parallel computations~\cite{Lima:2019:HEE,Trefethen:2013:EAS}, where the connection is significant, but a bit fuzzier. Furthermore, GPU utilization and GPU power also have a strong correlation (spearman's rho=0.81, pvalue=2.65e-08). In this case, however, the correlation is a bit less strong. This is somewhat puzzling because we know that GPU power depends directly on GPU utilization (due to voltage scaling). It suggests that the frameworks may not be utilizing the GPU as efficiently as they could and there is room for improving their efficiency. This is reinforced by the ratio between mean CPU energy and mean GPU energy across all the executions. The former corresponds to about 51.71\% of the latter. Even after removing a few outliers, the ratio is still 38.43\%. Considering that all the experiences are running on a commodity CPU, we can say that a lot of work is happening on the CPU.

\subsection{Energy efficiency of different frameworks}

To tackle RQ1, we initiate the analysis of ResNet and MobileNet with a batch size of 1. As depicted in Table 1, in MobileNet, TensorFlow demonstrates significantly slower inference time, approximately 6\textit{x} slower than PyTorch, and roughly 10\textit{x} slower than MXNet. Its total energy usage is 3.3\textit{x} higher than MXNet's for MobileNet and 2.4\textit{x} higher for ResNet. The differences are similar, albeit a bit lower, against PyTorch. The observed performance gap can be attributed to TensorFlow's suboptimal GPU utilization. Instead, it puts a significant portion of the workload on the CPU. This inefficiency results in prolonged inference time and increased total energy consumption. Conversely, MXNet stands out by effectively leveraging the GPU and spending comparatively less energy on the CPU. For ResNet, PyTorch's CPU energy usage is 68.8\% higher than MXNet's, which helps explain why the latter consumes less energy, overall.

\begin{table*}[bt]
    \centering
    \caption{Comparative results of inference of Resnet-50 with batch size 1 on CUDA and TensorRT execution providers}\label{tab:rq3}
    \resizebox{\textwidth}{!}{%
    \begin{tabular}{llrrrr>{\columncolor[HTML]{D3F3E7}}r>{\columncolor[HTML]{DAE8FC}}r}
    \hline
        Framework & Execution provider & \begin{tabular}[r]{@{}r@{}}GPU Utilization (\%)\\ (Cores, Memory)\end{tabular} & GPU power (W) & GPU Energy (J) & CPU Energy (J) & Total Energy (J) & \begin{tabular}[r]{@{}r@{}}Inference\\ Time (Sec)\end{tabular}
        \\ \hline
        mxnet & CUDA & (78.27, 24.12) & 40.97 & 2057.62 & 450.64 & 2508.26 & 50.22 \\ 
        ~ & TensorRT & (68.81, \textbf{15.38}) & 42.22 & \textbf{1530.31} & \textbf{389.02} & \textbf{1919.32} & \textbf{36.24} \\ \hline
        tf & CUDA & (67.88, 21.75) & 49.36 & 2326.07 & \textbf{416.02} & 2742.09 & \textbf{47.12} \\ 
        ~ & TensorRT & (55.68, \textbf{13.09}) & 39.64 & \textbf{1920.72} & 504.64 & \textbf{2425.36} & 48.45 \\ \hline
        torch & CUDA & (76.42, 23.81) & 51.63 & 2197.1 & 476.26 & 2673.36 & 42.56 \\ 
        ~ & TensorRT & (70.58, \textbf{20.3}) & 53.15 & \textbf{1703.48} & \textbf{410.64} & \textbf{2114.11} & \textbf{32.05} \\ \hline
    \end{tabular}
    }
\end{table*}

With a batch size of 64, TensorFlow shows an improvement. For example, it uses up only 35\% more total energy than MXNet for ResNet. However, overall it is still the least efficient of the three for ResNet and MobileNet. Notably, PyTorch and MXNet demonstrate comparable performance in this scenario. PyTorch makes more efficient use of the CPU, which leads to using overall 3.78\% less energy than MXNet for MobileNet and only 1.23\% more for ResNet.

In terms of accuracy (Table~\ref{tab:framework_accuracies}), MXNet closely competes with other frameworks in the case of MobileNet and even surpasses TensorFlow on ResNet. On the other hand, for vision applications where accuracy is a consideration, PyTorch attains the highest accuracy with a minimal impact on total energy consumption and duration.
The table only reports the accuracy of the original models (not converted models), as it did not change after conversion to ONNX, confirming the results of previous work~\cite{openja2022empirical}.

In contrast, for the BERT model, PyTorch outperforms both MXNet and TensorFlow. In this case, MXNet shows the poorest functionality in terms of both total energy and inference time when the batch size is 1. This stems mostly from very high CPU energy use. Besides, when taking accuracy into account, MXNet outshines its rivals, achieving around 11\% higher accuracy in the BERT model. PyTorch in this case exhibits the lowest accuracy, albeit only 0.5\% lower than TensorFlow's. These results highlight an important consideration for ML engineers and application developers: different frameworks perform better for different kinds of ML problems. For long-running applications and services that will serve a large number of users, experimenting previously with different options may save energy and improve accuracy in the long run. These findings reinforce Georgiou et al.'s conclusion in~\cite{georgiou2022green} that PyTorch is the less costly framework in terms of inference time and total energy than TensorFlow. While their findings suggest that the least costly framework always uses the least energy on both CPU and GPU, our results indicate that this statement may not hold true in all cases.

\subsection{Original vs. ONNX-Converted Models}
We proceed with our investigation in a similar manner as in RQ1 and further examine batch sizes of 1 and 64, this time focusing on exporting the ONNX format of each pre-trained model and running the inference of the exported model on the ONNX Runtime. Table ~\ref{tab:rq1} shows that the previously observed poor performance of TensorFlow at batch size 1 is mitigated after conversion, highlighting an optimization improvement achieved through ONNX. In batch size of 1, beyond TensorFlow, ONNX models not only outperform the original models by demonstrating an improvement in inference time across all cases but also contribute to a reduction in energy consumption. With a batch size of 64, in the ONNX versions converted from MXNet in both MobileNet and ResNet models, there is an observed around 13\% increase in inference time and approximately 10\% increase in total energy consumption. Similarly, the ONXX MobileNet model converted from PyTorch exhibits an increase of approximately 10\% in both energy consumption and inference time.
These results further reinforce the point made for RQ1: if inference time and energy efficiency are critical, experimentation is required because there is no overall winning solution, even when employing a runtime aimed at making inference efficient. 

It's noteworthy to mention that in all cases the use of ONNX brought an improvement in GPU utilization, with remarkable results observed in the BERT model implemented in PyTorch. It reaches around 100\% utilization, consuming less than half the energy of the original model for batch size 1 and 38.66\% less for 64. 

\subsection{Execution providers: CUDA vs. TensorRT}

Table~\ref{tab:rq3} presents results of running ResNet for each framework using two different execution providers, CUDA and TensorRT. Results for CUDA are a bit different from the ones reported in Table~\ref{tab:rq1} because of the methodological differences mentioned in Section~\ref{sec:meth}.
The table reveals that TensorRT is consistently more efficient than CUDA. It utilizes the GPU more effectively, leading to lower GPU energy usage and an overall reduction in total energy consumption. At the same time, it uses consistently less GPU memory. Furthermore, apart from TensorFlow, which experiences a marginal 2\% increase in inference time, both MXNet and PyTorch exhibit significant reductions of 27\% and 24.6\% in inference time, respectively. 

\section{Concluding remarks}
We conducted a preliminary study on the impact of DL runtime infrastructure on the energy efficiency of inference. The main takeaway message of this study is that there is no overall winner. Frameworks perform differently considering varying models, batch sizes, and runtime environments. For example, the original MXNet models outperformed both TensorFlow and PyTorch for the vision models and batch size 1. Notwithstanding, even though it is considered the \textit{``DL framework of choice at AWS''}~\cite{Vogels:2016:MDL} due to it being \textit{``the most scalable framework''}, the converted MXNet ONNX model exhibited higher energy usage and inference time for batch size 64. Furthermore, for the BERT model, it was significantly outperformed by PyTorch. We did obtain consistent results when using TensorRT as the execution provider for ONNX, though; it exhibited consistently lower inference time and energy consumption than CUDA.

Our results highlight that there is still a lot of room for investigating the energy impact of DL runtime infrastructures. So far, we have only investigated how different execution providers affect a single model, ResNet. In the future, we intend to extend this analysis to other models. Furthermore, ONNX models may be invoked using different client languages, such as Java, Python, C++, Rust, and JavaScript. In our preliminary (not included in the paper) examination, different languages seem to impose a non-negligible overhead. We plan to study this in the future. Another thread for future work is analyzing the impact of optimizations and runtime infrastructure execution settings. Previous work in the domain of compiler optimizations has shown that they are mercurial in nature~\cite{Mytkowicz:2009:PWD}. We are not aware of previous work investigating these optimizations, with a focus on energy, in the context of DL runtime infrastructures. Finally, we would like to investigate the energy behavior of at least one (more recent) large language model when employed to generate text.

\section{Acknowledgments}
We would like to thank the anonymous reviewers who provided invaluable suggestions to improve this paper. This research was partially supported by INES 2.0 (FACEPE PRONEX APQ 0388-1.03/14 and APQ-0399-1.03/17, CNPq 465614/2014-0).
\balance
\newpage
\bibliographystyle{ACM-Reference-Format}
\bibliography{references}
\end{document}